\documentclass[epj]{svjour}
\usepackage{epsfig}
\onecolumn

\newcommand{\be}{\begin{equation}}
\newcommand{\ee}{\end{equation}}
\newcommand{\vek}{\mbox{\boldmath${\rm k}$}}
\newcommand{\veq}{\mbox{\boldmath${\rm q}$}}

\begin{document}
%\hspace{9.8 cm}FZJ--IKP(TH)--2003--14

\title{The radiative decays $\phi\to\gamma a_0/f_0$ in the molecular
model for the scalar mesons}
\author{Yu. S. Kalashnikova$^1$, A. E. Kudryavtsev$^1$, A. V. Nefediev$^1$, 
C. Hanhart$^2$, J. Haidenbauer$^2$}
\institute{
Institute of Theoretical and Experimental Physics,\\
117259, B. Cheremushkinskaya 25, Moscow, Russia 
\and
Institut f\"ur Kernphysik (Theorie), Forschungszentrum J\"ulich,
D-52425 J\"ulich, Germany
}

\date{Received: date / Revised version: date}

\abstract{
We investigate the radiative decays of the $\phi$ meson to the scalar mesons
$a_0(980)$ and $f_0(980)$.
We demonstrate that, contrary to earlier claims, these decays should be of the 
same order of magnitude for a molecular state and for a compact state and, 
therefore, the available experimental information is consistent with both, 
a molecular as well as a compact structure of the scalars. 
Thus, the radiative decays of the $\phi$ meson into scalars
establish a sizable $K\bar K$ component of the scalar mesons, but do not
allow to discriminate between molecules and compact states.
}
\maketitle

\section{Introduction}

It has been claimed for many years that studies of radiative decays $\phi 
\to \gamma a_0(980) \to \gamma \pi^0 \eta$ and $\phi \to \gamma f_0(980) 
\to \gamma \pi^0 \pi^0$ are a powerful tool to discriminate between 
various models for the low-lying scalar mesons. The extraction of the
$\phi \gamma a_0$ and $\phi \gamma f_0$ coupling constants from the data is not a 
straightforward task (see \cite{Pennington}), 
but it is a common belief that, with data accurate enough, radiative 
decays would reveal the nature of the lightest scalars. 
   
The simplest mechanism for 
these radiative decays assumes that the $a_0$ and $f_0$ are $^3P_0$ 
quarkonia, and the decays proceed via a quark loop. Nevertheless, 
with the $\phi$-meson being mostly an $s \bar s$ state, this mechanism cannot 
be responsible for the decay $\phi \to \gamma a_0$, since, in the 
quarkonium picture, the $a_0$ is an isovector state made of light quarks.
Similarly, only $f_0(s \bar s)$ can be produced via the quark loop 
mechanism and, if so, the subsequent decay $f_0 \to \pi^0\pi^0$ is 
suppressed by the OZI rule. On the other hand, as both $f_0$ and $a_0$ are 
close to the $K \bar K$ threshold and are known to couple strongly to this 
channel, one expects that the radiative decay mechanism via charged kaon 
loop should play an essential role, as it was suggested in Refs.~\cite{AI,CIK,AGS}. 
The existing data on $\phi$ radiative decays \cite{SND,CMD,KLOE} support 
this expectation, as is shown in detail in Ref.~\cite{Achasov}.  

The latter observation does not mean {\it per se} that the quarkonium 
assignment for $a_0$ and $f_0$ is excluded by the data. It only means 
that the strong coupling to the $K \bar K$ channel, together with the threshold 
enhancement phenomenon, makes the kaon loop mechanism dominant. However, the 
strong coupling to $K \bar K$ implies that the $K \bar K$ 
component in the wave functions of these mesons should be large, and recent 
studies \cite{W} based on the analysis of near-threshold data confirm 
this. A large $K \bar K$ admixture should 
be reflected somehow in the radiative decay amplitude. 

In Ref.~\cite{CIK} it is claimed that there should be a strong suppression of
the $\phi \to \gamma f_0/a_0$ branching ratio for the scalars in case they are
loosely bound molecules as compared to pointlike scalars that correspond to
compact quark states, ($10^{-5}$ {\it vs} $10^{-4}$).  A study by Achasov et
al. \cite{AGS}, where the finite width of scalars was taken into account,
arrived at the same conclusion.  Thus, the authors of \cite{CIK} and
\cite{AGS} stress that data for this branching ratio should allow to prove or
rule out the molecular model of the scalars. However, no such suppression was
found in recent kaon loop calculations, Refs.~\cite{Oset,Markushin,Oller},
where the scalars were considered as dynamically generated states, {\em i.e.},
as molecules. The aim of the present paper is to demonstrate explicitly the
implications of a molecular structure of scalars on the radiative $\phi$
decay. In the course of this we can demonstrate what went wrong in the
analysis of Ref.~\cite{CIK} and confirm the results of
Refs.~\cite{Oset,Markushin,Oller}.

\section{Point--like scalars}

To simplify the situation we work with stable scalars --- the generalization to
a more realistic case is straight forward and should not change the
conclusions; we comment on what is necessary for this generalization in what follows.
The current describing the radiative transition between the vector meson $\phi$ and a scalar meson $S$, 
in the kaon loop model, is written as \cite{Nussinov,Lucio} (see \cite{CIK} for notations)
\be
M_{\nu}=e\frac{g_{\phi}g_S}{2\pi^2im^2_K}I(a,b)[\varepsilon_{\nu}(p\cdot q)-p_{\nu} (q\cdot \varepsilon)],
\label{loop}
\ee
where $p$ and $q$ are the momenta of the $\phi$-meson and the photon, respectively, $m_K$ is the kaon mass,
$g_\phi$ and $g_S$ are the $\phi K^+K^-$ and $SK^+K^-$ coupling constants, 
$\varepsilon_\nu$ is the polarization four-vector of the $\phi$-meson,
$a=\frac{m^2_\phi}{m^2_K}$, and $b=\frac{m^2_S}{m^2_K}$ (in case of an
unstable particle produced $m^2_S$ is to be replaced by the invariant mass
squared of the decay products). 
The amplitude (\ref{loop}) is transverse, $M_{\nu}q_{\nu}=0$, and is proportional to the photon momentum.
 
For the pointlike model of the scalar mesons
the function $I(a,b)$ was calculated in Refs.~\cite{AI,CIK}.
It is given by 
\be
I(a,b)=\frac{1}{2(a-b)}-\frac{2}{(a-b)^2} 
\left[f\left(\frac{1}{b}\right)-f\left(\frac{1}{a}\right)\right]
+\frac{a}{(a-b)^2}\left[g\left(\frac{1}{b}\right)-g\left(\frac{1}{a}\right)\right],
\label{I}
\ee
$$
f(\alpha)=\left\{
\begin{array}{ccc}
-[\arcsin(\frac{1}{2\sqrt{\alpha}})]^2,&\quad&\alpha>\frac{1}{4}\\
\frac{1}{4}\left[\ln(\frac{\eta_+}{\eta_-})-i\pi \right],&&\alpha<\frac{1}{4}
\end{array}
\right.
$$
$$
g(\alpha)=
\left\{
\begin{array}{ccc}
\sqrt{4\alpha-1}\arcsin(\frac{1}{2\sqrt{\alpha}}),&\quad&\alpha>\frac{1}{4}\\
\frac{1}{2}\sqrt{1-4\alpha}\left[\ln(\frac{\eta_+}{\eta_-})-i\pi\right],&&\alpha<\frac{1}{4}
\end{array}
\right.
$$
$$
\eta_{\pm}=\frac{1}{2\alpha}\left(1\pm\sqrt{1-4\alpha}\right).
$$
Note that the integral $I(a,b)$ remains finite in the limit $a \to b$.

To arrive at the formula (\ref{loop}) consider the sum of the graphs depicted in 
Fig.~1(a)-(c), where the appearance of the graph 1(c) is a consequence of gauge
invariance, since the $\phi\to K\bar K$ vertex is momentum-dependent. 
The current in Eq. (\ref{loop}) is given
by $M_{\nu}=eg_{\phi}g_S\varepsilon_{\mu}J_{\mu\nu}$,
with 
\be
J_{\mu \nu}=J^{(a)}_{\mu \nu}+J^{(b)}_{\mu \nu}+J^{(c)}_{\mu \nu}=2J^{(a)}_{\mu \nu}+J^{(c)}_{\mu \nu},
\label{sum}
\ee
where 
\be
J^{(a)}_{\mu \nu}=\int 
\frac{d^4k}{(2\pi)^{4}}\frac{(2k-p)_{\mu}(2k-q)_{\nu}}
{[k^2-m^2+i0][(k-q)^2-m^2+i0][(k-p)^2-m^2+i0]},
\label{a}
\ee
\be
J^{(c)}_{\mu \nu}=-2g_{\mu \nu}\int 
\frac{d^4k}{(2\pi)^4}\frac{1}{[k^2-m^2+i0][(q+k-p)^2-m^2+i0]},
\label{c}
\ee
and $m=m_K$.

\begin{figure}[t]
\begin{center}
\begin{tabular}{cccc}
\epsfig{file=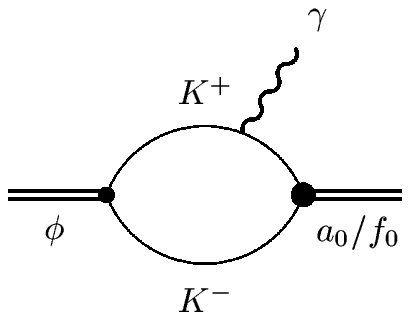,width=3.5cm}&
\raisebox{-6mm}{$\epsfig{file=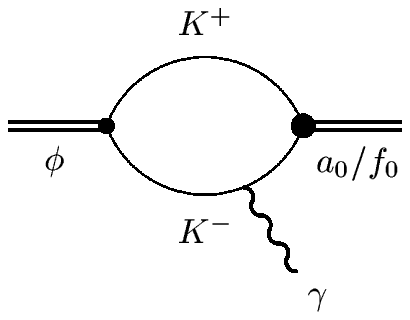,width=3.5cm}$}&
\epsfig{file=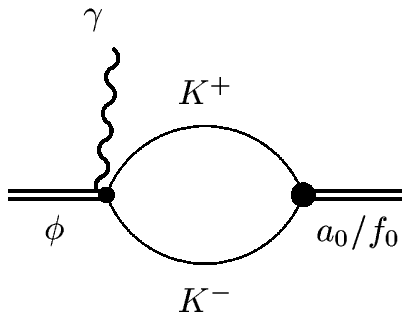,width=3.5cm}&
\epsfig{file=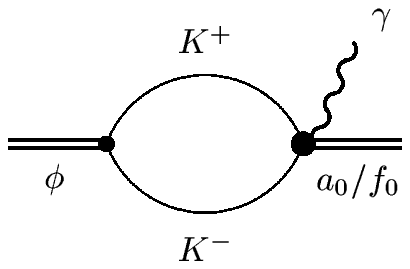,width=3.5cm}\\
(a)&(b)&(c)&(d)
\end{tabular}
\end{center}
\caption{Diagrams contributing to the radiative decay amplitude (\ref{loop}).}
\end{figure}

Since gauge invariance demands the structure of the integral (\ref{sum}) to be
\be
J_{\mu \nu}=J[p_{\nu}q_{\mu}-(p \cdot q)g_{\mu \nu}],
\label{structure}
\ee 
the strategy applied in Ref. \cite{CIK} is to read off the coefficient of the 
$p_{\nu}q_{\mu}$ term, coming entirely from the integral
(\ref{a}), and to restore then the coefficient of the $g_{\mu \nu}$ term with the 
help of Eq.~(\ref{structure}). This allows the authors to deal with a finite integral 
and thus to bypass the problem of 
treating the divergent parts of the loop integrals (\ref{a}), (\ref{c}). 
However, as we shall see below, the divergent pieces cancel and the sum of 
diagrams given in Eq.~(\ref{sum}) is finite \cite{Bramon}.

To see this we decompose the expression for $J_{\mu\nu}^{(a)}$ as
\be
2J^{(a)}_{\mu\nu}=J[p_{\nu}q_{\mu}-(p \cdot q)g_{\mu \nu}]+2g_{\mu \nu}J'_a \ ,
\label{tloop}
\ee
where
\be
J=-\frac{i}{2\pi^2m^2}\left\{\frac{1}{(a-b)}\int^1_0 dz 
\left[1-z-\frac{1-az(1-z)}{z(a-b)}\ln\frac{1-bz(1-z)}{1-az(1-z)}\right]\right.
\label{Ipq}
\ee
$$
\hspace*{3cm}\left.-\frac{i\pi}{(a-b)^2}\int^{1/\eta_+}_{1/\eta_-} 
dz\left[\frac{1}{z}-(1-z)a\right]\right\}=-\frac{i}{\pi^2m^2}I(a,b).
$$
Here and in what follows we consider the case of $m_{\phi}>2m$, $m_S<2m$. In addition 
\be
J'_a=\frac{i}{16\pi^2}\left[\frac{2}{\varepsilon}-\gamma_E-\ln\frac{m^2}{4\pi\mu_\varepsilon^2}\right]
-\frac{i}{8\pi^2}\int^1_0 dz (1-z)\ln [1-bz(1-z)], \ee where $\mu_\varepsilon$
is the auxiliary mass parameter, the number of dimensions $D$ is equal to
$4-\varepsilon$, and $\gamma_E$ is the Euler constant. Similarly, the contact
term (\ref{c}) can be presented as $-2g_{\mu \nu}J'_c$ with \be
J'_c=\frac{i}{16\pi^2}\left[\frac{2}{\varepsilon}-\gamma_E-\ln\frac{m^2}{4\pi\mu_\varepsilon^2}\right]-
\frac{i}{16\pi^2}\int^1_0 dz \ln [1-bz(1-z)],
\label{contact}
\ee
and, since
$$
\int^1_0 dz (1-2z)\ln[1-bz(1-z)]=0,
$$
the structure (\ref{structure}) is restored. We conclude therefore that, with 
the proper regularization, the total matrix element is finite. It means that the range 
of convergence of the integrals involved 
is defined only by the kinematics of the problem. In particular, if both masses of 
the vector and scalar mesons are close to the $K \bar K$ threshold, the integrals 
converge at $k_0\sim m$ and for nonrelativistic values of the three-dimensional loop momentum $\vek$, 
$|\vek| \ll m$. The nonrelativistic limit of the integral $I(a,b)$ takes the form
\be
I_{NR}(a,b)=\frac{\pi(x^3+3xy^2)}{24(x^2+y^2)^2}+i\frac{\pi y^3}{12(x^2+y^2)^2},
\label{nonrel}
\ee  
where 
$$
y=\sqrt{(a/4)-1} \, \quad x=\sqrt{1-(b/4)} \, ,\quad x,y\ll 1.
$$
Note that, although the 
expression (\ref{nonrel}) contains the factor $\frac{1}{x^2+y^2} 
\sim\frac{1}{a-b}$, it does not mean that $I_{NR}(a,b)$ blows up in the limit 
of zero photon energy, $\omega\rightarrow 0$. Indeed, the formula (\ref{nonrel}) is valid for the scalar 
meson lying below the $K \bar K$ threshold, so one cannot put $\omega=0$ here. If 
the scalar appears above the kaon threshold, Eq.~(\ref{nonrel}) is replaced by
\be
\frac{\pi i}{24}\frac{(2y+\tilde{x})}{(y+\tilde{x})^2},\quad \tilde{x}=\sqrt{(b/4)-1}
\label{above}
\ee
so that $I(a,b)$ remains finite in the limit $\omega\to 0$.

\section{Introducing the scalar wave function}

When treating the scalar meson as an extended (non-pointlike) object it is not
sufficient to insert the corresponding form factor into the $K^+ K^- S$ vertex
(see \cite{CIK}), but gauge invariance calls for a correction term induced by
this additional flow of charge. Since only soft photons are involved the
needed correction term can be expressed as the derivative of the formfactor inserted.
Thus we get for the induced vertex
\be
\Gamma_{\nu}(K ^+ K^- S\gamma)=-2(p^+_{\nu}-p^-_{\nu})\left. \frac{\partial 
\Gamma(p^2,m^2)}{\partial p^2}\right|_{p^2=m^2},
\label{scalar}
\ee   
where $\Gamma(p_+^2,p_-^2)=\Gamma(p_-^2,p_+^2)$ parameterizes the momentum 
dependence of the $K^+K^-S$ 
vertex, with $\Gamma (m^2,m^2)=1$. Here 
$p^+_{\nu}$ and $p^-_{\nu}$ are the $K^+$ and $K^-$ four-momenta, respectively.
The corresponding extra diagram is depicted at Fig~1(d).

Before proceeding further we note that inclusion of the extra contact vertex 
(\ref{scalar}) is a way to insert an ultraviolet cutoff in a 
gauge invariant way. As demonstrated above, the integrals of interest converge already for
nonrelativistic momenta even for a pointlike vertex, thus it is justified
to use nonrelativistic kinematics also when the vertex 
function $\Gamma$ is included, as it was 
done in \cite{Markushin} --- one only needs the mild assumption that 
$\Gamma$ decreases faster than $1/k$ for increasing values of its 
arguments. 
Then only the positive-energy parts of the kaon propagators are retained, the kaon energies are replaced by $m$,  
and $m_\phi$ and $m_S$ are replaced by $2m$, wherever possible. 
As to the vertex function, in the nonrelativistic description the 
virtuality of kaons is measured by the relative momentum of kaons in the 
intermediate state, so that in the 
center-of-mass frame of the vector meson ($\vec{p}=0$) the vertex 
function $\Gamma$ is a function of the three-momentum of the 
outgoing kaons only and thus the spatial loop integrals read
\be
J_{ik}=
2J^{(a)}_{ik}+J^{(c)}_{ik}+J^{(d)}_{ik}=-\delta_{ik}\frac{i}{4\pi^2}(a-b)
I(a,b;\Gamma)+...~,
\label{tot}
\ee
when evaluated in the rest frame of the vector meson. Terms that do not
contribute to the process of interest are not shown explicitly.
Note, gauge invariance is enshured by the appearance of the term $(a-b)$ that
vanishes for vanishing outgoing photon energy.
 The individual
integrals are
\begin{eqnarray}
\nonumber
2J^{(a)}_{ik}&=&-\frac{i}{m^3}\int\frac{d^3 k}{(2\pi)^3}\frac{k_ik_k\Gamma(|\vek-\veq/2|)}{[E_V-\frac{k^2}{m}+i0]
[E_S-\frac{(\vek-\veq/2)^2}{m}+i0]}, \\
\nonumber
J^{(c)}_{ik}&=&-\frac{i}{2m^2}\delta_{ik}\int\frac{d^3 k}{(2\pi)^3}\frac{\Gamma(k)}{E_S-\frac{k^2}{m}+i0},
\\
J^{(d)}_{ik}&=&-\frac{i}{2m^2}\int 
\frac{d^3k}{(2\pi)^3}\frac{k_ik_k}{E_V-\frac{k^2}{m}+i0}\frac{1}{k}
\frac{\partial \Gamma(k)}{\partial k},
\label{explint}
\end{eqnarray}
We assume $E_V=m_V-2m>0$, $E_S=m_S-2m<0$ for looking at only one kinematic
regime is sufficient to make our point clear. For more realistic calculations
that include the finite width of the scalar mesons we recommend Refs.
\cite{Oset,Markushin,Oller}.  Performing integration by parts in the integral
$J^{(d)}_{ik}$, one has 
\be J^{(d)}_{ik}= \frac{i}{2m^2}\delta_{ik} \int
\frac{d^3 k}{(2\pi)^3} \frac{\Gamma(k)}{E_V-\frac{k^2}{m}+i0}+
\frac{i}{3m^3}\delta_{ik} \int \frac{d^3 k}{(2\pi)^3}
\frac{k^2\Gamma(k)}{(E_V-\frac{k^2}{m}+i0)^2}.
\label{byparts}
\ee
This trick was used both in Refs.~\cite{CIK} and \cite{Markushin}. 

Let us now assume that $\Gamma$ decreases with the range $\beta$ that 
satisfies the conditions
\be
\beta^2 \gg mE_V,\quad\beta^2 \gg m|E_S|.
\label{beta}
\ee
With the help of the representation (\ref{byparts})
one immediately sees  
that, in the limit $\beta \rightarrow \infty$, the
divergent terms in $J_{ik}$, Eq.~(\ref{tot}), cancel each other and, in the leading 
nonrelativistic approximation, $E_V \ll m$, $|E_S| \ll m$, the total matrix 
element does not depend on $\beta$:
\be
I(a,b;\Gamma)=I_{NR}(a,b).
\label{nonreltot}
\ee

We stress that the result (\ref{nonreltot}) follows from the 
nonrelativistic formula (\ref{tot}), and the only condition needed is 
(\ref{beta}). 

We have repeated the calculation of $I(a,b;\Gamma)$ presented in 
Ref.~\cite{Markushin} with the model form factor $\Gamma(\vek)=\beta^2/(\vek^2+\beta^2)$.
The results are depicted at Fig.~2 together with the results of the full 
pointlike theory. One can see that, in the soft-photon limit, there is no
considerable suppression of the matrix element due to finite values of 
$\beta$, down to $\beta \sim 0.3$ GeV. The reason for this was discussed 
above --- the integral of Eq. (\ref{sum}) converges for
nonrelativistic values of $|\vek|$, in the soft-photon limit.

\begin{figure}[t]
\centerline{\epsfig{file=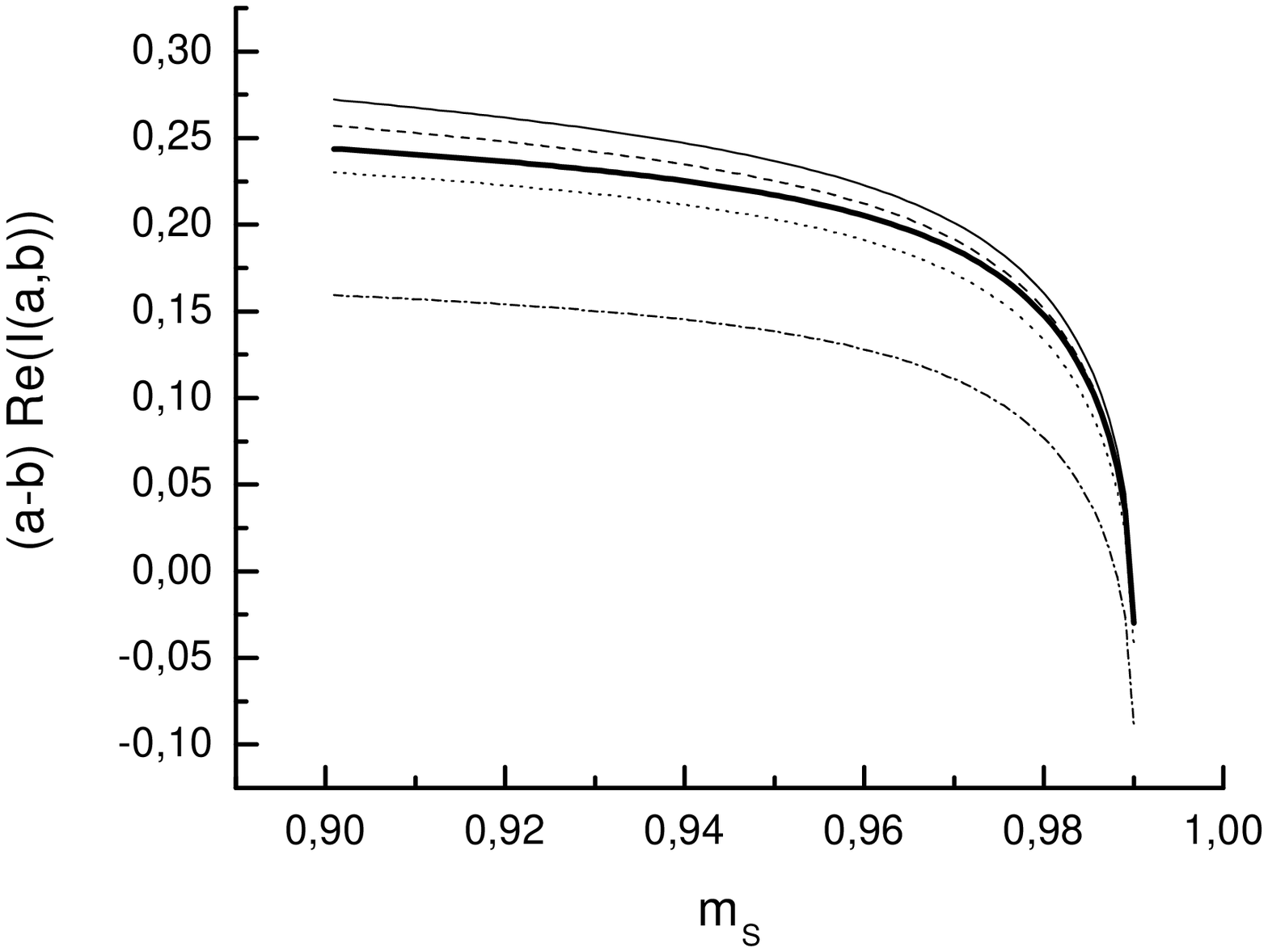,width=8cm}\hspace*{3mm}\epsfig{file=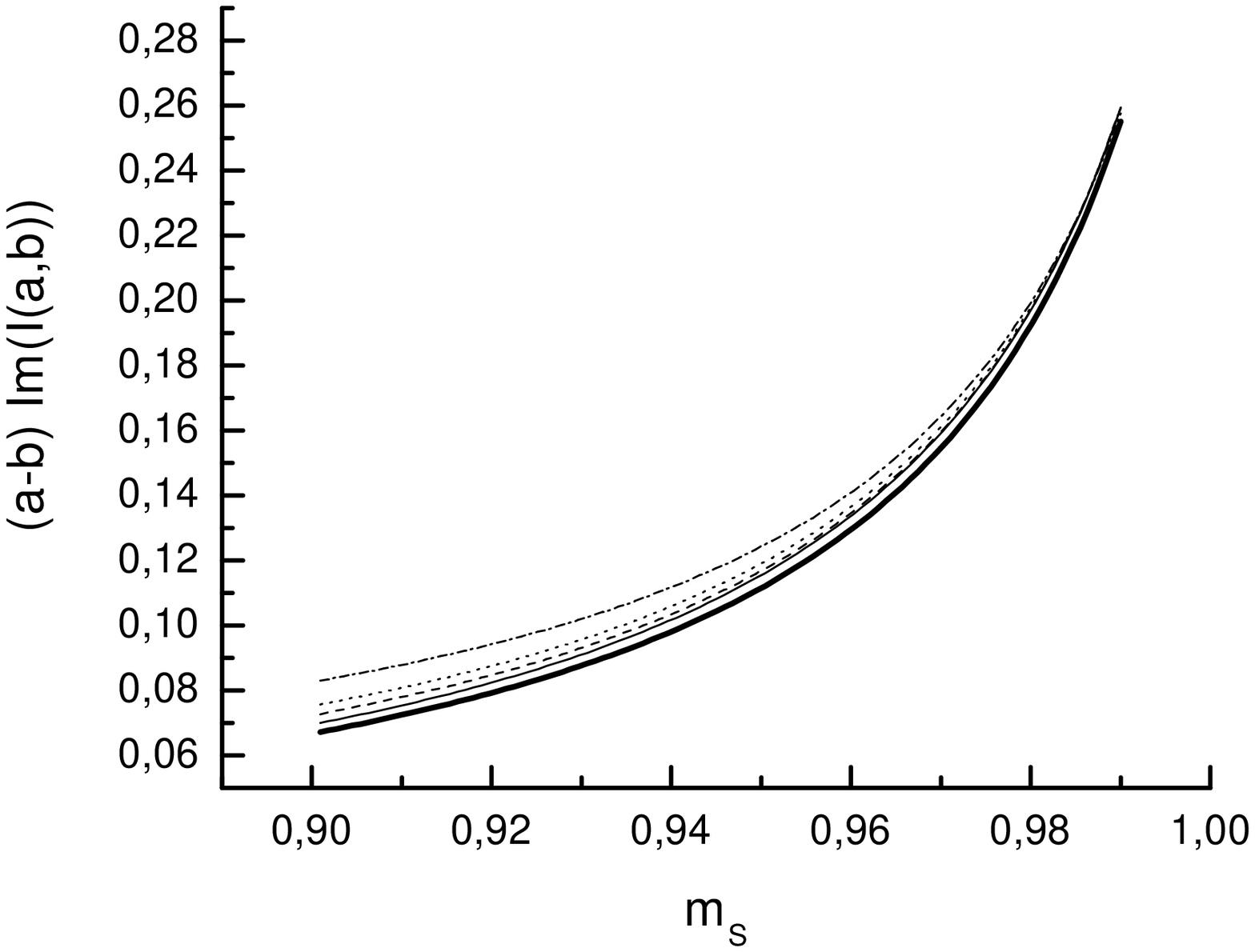,width=8cm}}
\caption{The real (the left plot) and the imaginary (the right plot) parts of 
the function $I_1=(a-b)I(a,b;\Gamma)$
for $\beta=0.2$ GeV (dash-dotted line), $\beta=0.4$ GeV (dotted line), $\beta=0.6$ GeV (dashed line), and
$\beta=0.8$ GeV (thin solid line). The result of the full pointlike theory is given by the thick solid line.}
\end{figure}

Now we specify the form factor in the molecular model for the scalar mesons. To this 
end we use the well-known quantum--mechanical expressions which relate the 
$K\bar{K}S$ vertex and the wave function of the molecule. In the vicinity 
of a bound state the nonrelativistic $t$-matrix $t(\vek,\vek',E)$ takes the form
\be
t(\vek,\vek',E)=\frac{\gamma(\vek)\gamma(\vek')}{E+\varepsilon-i0},\quad 
\gamma(\vek)=\hat{v}\phi(\vek),
\label{t}
\ee 
where $\phi(\vek)$ is the bound--state wave function in the momentum 
space,
normalized to unity, $\varepsilon=-E_S$ is the binding energy, and the
Schr{\" o}dinger equation for the bound state is written symbolically as
\be
\frac{\vek^2}{m}\phi(\vek)+\hat{v}\phi(\vek)=-\varepsilon\phi(\vek).
\label{eq}
\ee
The relativistic vertex differs from the nonrelativistic vertex $\gamma$ 
by a kinematical factor (see, {\em e.g.}, \cite{Landau}),
\be
g_S\Gamma(\vek)=(2\pi)^{3/2}\sqrt{8m^2m_S}\;\gamma(\vek),
\ee
where the effective coupling $g_S$ is introduced
to ensure the normalization condition $\Gamma(0)=1$. 
Using the bound--state equation (\ref{eq}), one has, finally,
\be
g_S\Gamma(\vek)=(2\pi)^{3/2}\sqrt{8m^2m_S}\left(\frac{\vek^2}{m}+\varepsilon\right)\phi(\vek).
\label{vertex}
\ee 
Thus we find that the momentum dependent factor that appears in
Eq. (\ref{vertex}) exactly compensates for the two kaon propagator in
Eq. (\ref{explint}). The wavefunction then supplies exactly that piece due to
its demanded asymptotics.

A real molecule is a loosely bound state with a large mean 
distance between the constituents --- much larger than the range of the binding force $r_0$.
In this deuteronlike case one has
\be
\phi(\vek)=\frac{\sqrt{\kappa}}{\pi}\frac{1}{\vek^2+\kappa^2},\quad\kappa=\sqrt{m\varepsilon}.
\label{deuteron}
\ee
Correspondingly, the vertex (\ref{vertex}) does not depend on $\vek$, and one 
can safely use the formulae (\ref{loop}), (\ref{I}) of the pointlike 
theory with
\be
g_S=\frac{(2\pi)^{3/2}}{\pi}\sqrt{8m_S\kappa},\quad\frac{g_S^2}{4\pi}\approx 32m\sqrt{m\varepsilon}.
\label{constant}
\ee
The nonrelativistic expansion (\ref{nonrel}) of the integral $I(a,b)$ can 
be used as well.

So we conclude that the range $\beta$ of the form factor should be 
identified with the inverse range of the force, $\beta \sim 1/r_0$, and, if 
the inequality
\be
\kappa r_0 \sim \frac{\kappa}{\beta} \ll 1
\label{inequality}
\ee 
holds true, the results of the pointlike theory for the radiative $\phi \to 
\gamma S$ decay are valid for molecular model of the scalar. In particular, 
there is no special suppression of the matrix element due to a finite 
value of $\beta$.

The latter statement is based on the validity of the inequality 
(\ref{inequality}). What values of $\beta$ would one expect in realistic 
models of the $K \bar K$ molecule? In the meson-exchange models like 
\cite{Jue} it is argued that a strong $t$-channel force is responsible for the formation of scalars.
In such a case it is reasonable to identify $\beta$ with the mass of 
the lightest meson exchanged. As there is no pion exchange in the scalar 
sector, the lightest meson should be the $\rho$, which gives for  
$\beta$ the value of about $0.8$ GeV. In the phenomenological picture of 
Ref.~\cite{Markushin}, $\beta$ is taken to be $0.5\div 0.7$ GeV. In the quark language,
$\beta$ is defined by the scale of the internal size of the quark wave 
function, which also leads to the estimate for $\beta$ to be of the order 
of a few hundred MeV. With such estimates, the inequality 
(\ref{inequality}) is safely valid for the masses of the scalar about 
$970\div 980$ MeV.

The formula (\ref{constant}) implies that the vertex $g_S$ depends on the 
binding energy and its value decreases with decreasing binding energy.
This in turn causes a suppression of the branching ratio when the 
binding energy tends to zero, cf. Fig.~3. However, for binding energies
of typical order of magnitude, for example, $\varepsilon =10$ MeV,
Eq.~(\ref{constant}) yields a coupling constant $g_S$ of 
\be
\frac{g_S^2}{4\pi}=1.12 \ GeV^2 .
\label{numconst}
\ee
That corresponds to a branching ratio of 
$Br(\phi\to \gamma S) \approx 2.6 \times 10^{-4}$ which means that
there is practically no suppression. 

Nevertheless, we should emphasize in this context that a reliable quantitative 
calculation of the width certainly requires a more realistic approach
where it is taken into account that the scalar mesons have finite widths due to the 
presence of the light pseudoscalar channels, and that the quantities that 
are really measured are the transitions $\phi \to \gamma \pi\pi$ 
or $\phi \to \gamma \pi \eta$. 
The impact of finite width effects have been thoroughly investigated 
by J.A. Oller \cite{Oller} and also by Achasov and Gubin \cite{Gubin}
and we refer to their work for details. Here we only want to make the reader
aware of the fact that due to the proximity of the $\gamma S$ threshold to the 
mass of the $\phi (1020)$ resonance even small variations in the nominal 
resonance masses of the scalar mesons have a drastic effect on the available 
phase space and in turn on the obtained results -- 
as it can be imagined from  Fig.~3 -- 
unless the finite width of the ($f_0(980)$ or $a_0(980)$) scalar mesons is 
considered \cite{Oller,Gubin}. 

To take into account finite width effects 
one has to use the two-channel version of Eq.~(\ref{t}) from the very
beginning so
that the vertex which appears in the loop integral is accompanied by the
vertex that appears in the resonance decay, as it is required by the
two-channel unitarity condition. If the
characteristic scale $\beta$ in this full $t$-matrix is not too small,
then the feature that there is no specific suppression due to the
molecular structure of the scalar mesons will be preserved,
cf. appendix A.    

\begin{figure}[t]
\centerline{\epsfig{file=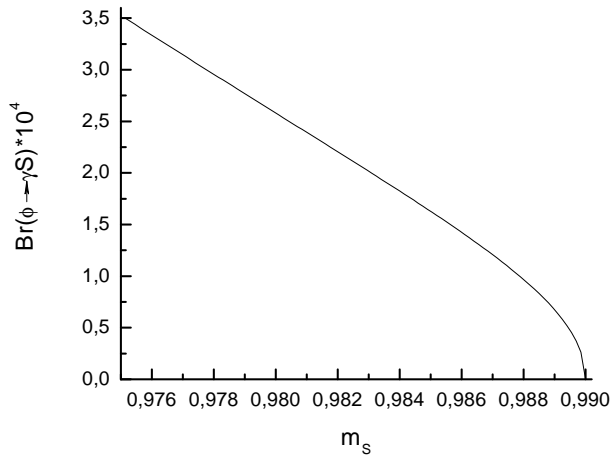,width=8cm}}
\caption{The dependence of the branching $Br(\phi\to\gamma S)$ on the mass 
of the scalar meson in the molecular model.}
\end{figure}

\section{Comparison to older work and conclusions}

Our findings are in contradiction with the results of Ref.~\cite{CIK}.  
The specific model for the $K \bar K$ 
molecule used there was taken from Ref.~\cite{Barnes}, which, in turn, is a 
modification of the approach developed in Ref.~\cite{Isgur} and based on the
quark--exchange picture. The $K \bar K$ interaction employed in Ref.~\cite{Barnes} 
was approximated by a local potential of the form
\be
V(r)=-V_0\exp{\left[-\frac{1}{2}\left(\frac{r}{r_0}\right)^2\right]},
\label{barnes}
\ee      
with $r_0=0.57$ fm. This interaction gives $\varepsilon=10$ MeV, so that
$\kappa r_0 \sim 0.2$, and the molecule is rather deuteronlike.

The wave function was parameterized as
\be
\psi(r)=\left(\frac{\mu^3}{\pi}\right)^{1/2}e^{-\mu r},\quad
\phi(\vek)=\frac{(2\mu)^{3/2}}{\pi}\frac{\mu}{(\vek^2+\mu^2)^2},
\label{psi}
\ee
with $\mu=0.144$ GeV. This wave function yields a good approximation for 
the exact wave function, in the momentum space (see \cite{CIK}). 
On the other hand, the wave function (\ref{deuteron}) 
with $\varepsilon=10$ MeV looks very similar, see Fig.~4. 

\begin{figure}[t]
\centerline{\epsfig{file=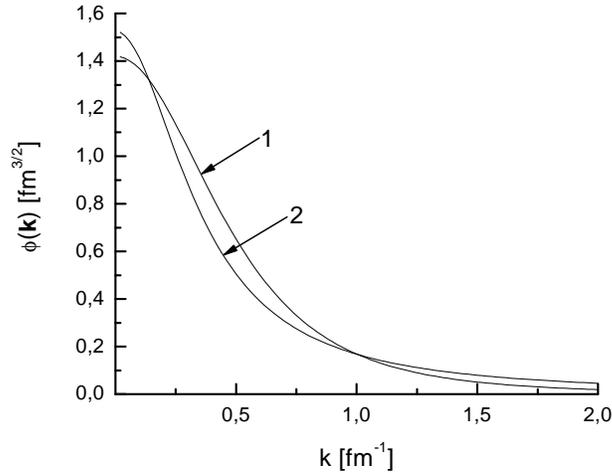,width=8cm}}
\caption{The wave function of the $K\bar K$ system, in momentum space. The approximate
solution Eq.~(\ref{psi}) --- the curve 1, 
and the deuteronlike wave function, Eq.~(\ref{deuteron}) --- the curve 2.}
\end{figure}

So what is wrong with Ref.~\cite{CIK}, and where does the suppression of the radiative 
decay amplitude come from? The answer is rather simple. In Ref.~\cite{CIK}, the calculations 
of the loop integrals were performed with using the wave function
\be
\phi(\vek)=\phi(0)\frac{\mu^4}{(\vek^2+\mu^2)^2},
\ee
as a form factor, instead of the correct formula (\ref{vertex}) for the 
form factor. This led to the result of $4\times10^{-5}$ for the branching
ratio (or $\Gamma(\phi\to\gamma S) = 1.7\times10^{-4}$ MeV). 
The same incorrect choice for the form factor was made in \cite{AGS}. As 
$\mu$ is as small as 0.144 GeV, no surprise that the suppression found was 
huge!  

The radiative decay width calculated with the parameterization of the wavefunction
(\ref{psi}) and the correct formula (\ref{vertex}) is 
$\Gamma(\phi\to\gamma S) = 2.4\times10^{-3}$ MeV. 
It is somewhat large as compared to the experimental result. 
We want to point out, however, that this is 
primarily due to the not very accurate parameterization. 
Indeed, the approximation (\ref{psi}) is definitely wrong for
distances beyond the range of the forces, $r \gg r_0$, where the wave function should 
behave as $\sqrt{\frac{\kappa}{2\pi}}\frac{e^{-\kappa r}}{r}$. On the other hand, 
the deuteronlike wave function is wrong at short distances. It is clear,
however, that possible contributions to the integral
(\ref{tot}) coming from short distances correspond to large values of 
$|\vek|$ where the integrand is suppressed. The value of $1.1\times 
10^{-3}$ MeV for the width, obtained in the pointlike theory with a
value of $g_S$ given by Eq.~(\ref{numconst}), is therefore a good
approximation to the corresponding width calculated within a molecular model \cite{Barnes}.

In conclusion, there is no considerable suppression of the $\phi\to\gamma S$ 
width in the molecular model for the scalar mesons. As soon as the form factors of
an extended scalar meson are treated properly, the corresponding results become
very similar to those for a pointlike scalar meson (quarkonium), provided reasonable values
are chosen for the range of the interaction. We confirm the range of order of $10^{-3}\div 10^{-4}$ 
for the branching ratio obtained in Refs.~\cite{Oset,Markushin,Oller}.

\begin{acknowledgement}
Instructive discussions with N.N. Achasov are acknowledged. 
This research is part of the EU Integrated Infrastructure Initiative
Hadron Physics Project under contract number RII3-CT-2004-506078, and was 
supported also by the DFG-RFBR grant no. 02-02-04001 (436 RUS 113/652).
Yu. S. K, A. E. K, and A. N. acknowledge the support of  
the Federal Programme of the Russian Ministry of Industry, Science, and Technology No 40.052.1.1.1112.
and of the grants NS-1774.2003.2 and RFBR 02-02-16465.
\end{acknowledgement}

\section{Appendix: Inclusion of a finite width}

In this appendix we discuss the effect of a finite width of the scalar mesons, 
due to their decay into two pseudoscalars ($P_1P_2$), on the total width 
$\Gamma(\phi\to\gamma S)$.

The $P_1P_2$ invariant mass distribution has the form 
\be
\frac{d\Gamma}{dm_{P_1P_2}}=\frac{\alpha g_\phi^2\omega}{3(2\pi)^6m_\phi^2}
|(a-b)I(a,b)|^2|A_{K\bar{K}\to P_1P_2}(m_{P_1P_2})|^2,\quad 
a=\left(\frac{m_\phi}{m}\right)^2,\quad b=\left(\frac{m_{P_1P_2}}{m}\right)^2,
\label{A1}
\ee
where $\omega=\frac{m_\phi^2-m_{P_1P_2}^2}{2m_\phi}$ is the photon energy and 
$m_{P_1P_2}$ 
is the invariant mass of the outgoing 
pseudoscalars. Here the range of the force is assumed to be
small enough so that one can take the integral $I(a,b)$ for the 
point-like case, cf. Eq.~(\ref{inequality}). 

To account for the finite width of the scalar mesons one is to use the 
two--channel $t$--matrix. For the deuteronlike case, the amplitude in 
the $K\bar K$ channel can be written in the scattering length approximation 
with a complex scattering length $a_{K \bar K}$,
\be 
a_{K \bar K}=\frac{1}{\kappa_1+i\kappa_2},~~\kappa_2>0,
\ee
for energies around the $K\bar K$ threshold (and energies sufficiently far
away from the $P_1P_2$ threshold). Then the $K\bar{K}\to P_1P_2$ transition 
amplitude $A$ squared can be found as
\be
|A_{K\bar{K}\to P_1P_2}(m_{P_1P_2})|^2=
\frac{64\pi^2m_\phi^2\kappa_2}{[\kappa_1-\sqrt{-mE}\Theta(-E)]^2+
[\kappa_2+\sqrt{mE}\Theta(E)]^2},
\label{A2}
\ee
with $E=m_{P_1P_2}-2m$. 

In the limit $\kappa_2 \to 0$ there is no coupling to the $P_1P_2$ channel
and, for $\kappa_1>0$, there is a bound state in the $K \bar K$ channel 
with the binding energy $\varepsilon = \kappa_1^2/m$. One can readily obtain 
the total radiative width in this case, which is given by the standard formula, 
\be
\Gamma(\phi\to\gamma S)=\frac{\alpha g_S^2 g_\phi^2\omega}{48\pi^4m_\phi^2}
|(a-b)I(a,b)|^2, \quad a=\left(\frac{m_\phi}{m}\right)^2,\quad 
b=\left(\frac{m_S}{m}\right)^2, \quad 
\omega=\frac{m_\phi^2-m_S^2}{2m_\phi},
\label{A4}
\ee
with $m_S=2m-\varepsilon$ and $g_S$ defined by Eq.~(\ref{constant}).
 
In order to estimate the effect of a finite inelasticity $\kappa_2$, we
have calculated the contribution to the total width,
\be
\Gamma_{tot} = \int dm_{P_1P_2} \frac{d\Gamma}{dm_{P_1P_2}}
\ee
 from the 
distribution (\ref{A1}) integrated over the near-threshold region, 
$900$ MeV $<M<m_{\phi}$. The results for the branching ratios are listed
in Table~1. One can see that 
the branching ratio remains in the order of $10^{-4}$ even for 
$\kappa_1=0$, if the scale of $\kappa_2$ is around $50\div 100$ MeV. We 
conclude, therefore, that the results presented in this paper are robust
against the inclusion of the finite width of the scalar.
  
We would like to note here that the above--mentioned scale for $\kappa_2$ is 
quite natural. For example, as it was shown in Ref.~\cite{Baru}, the 
data on the $\pi\pi$ scattering near the $K \bar K$ threshold
are, indeed, nicely described in the scattering length approximation, with
$\kappa_2$ lying in this range (and the ratio $\kappa_1/\kappa_2$ being of 
order unity).

\begin{table}[t]
\begin{center}
\begin{tabular}{|c|c|c|c|}
\hline
$\kappa_1$$\setminus$ $\kappa_2$& 0& 50 & 100\\
\hline
70&2.56&3.07&2.80\\
\hline
0&0&1.22&1.57\\
\hline
\end{tabular}
\end{center}
\caption{The branching ratio $Br(\phi\to\gamma S)\times 10^4$; $\kappa_{1,2}$ are given in MeV.}
\end{table}

\end{document}